\documentclass[letterpaper,11pt]{article}
\pdfoutput=1

 \usepackage{etex}
\usepackage[totalwidth=460pt, totalheight=680pt]{geometry}
\usepackage{setspace}

\usepackage{amsmath, array}
\usepackage{mathtools}
\usepackage{epstopdf}
\usepackage{slashed}
\usepackage{fancyvrb}
\usepackage{dsfont}

\usepackage[english,greek]{babel}
\usepackage{verbatim}
\usepackage{booktabs}
\usepackage{arydshln}

\usepackage[usenames,dvipsnames]{color}
\usepackage{color}
\usepackage{pstricks,framed}
\usepackage{tikz}
\usetikzlibrary{chains}

\usepackage{tensor}
\usepackage{amssymb}
\usepackage{amsthm}
\usepackage{simplewick}

\usepackage{chngcntr}

\usepackage[]{mciteplus}
\mciteSetBstSublistMode{n}
\mciteSetMidEndSepPunct{\quad$\bullet$\quad}{.}{\relax}
\usepackage{graphicx}
\usepackage[colorlinks=true, linktoc=page, citecolor=blue, urlcolor=blue]{hyperref}
\usepackage{mathdots}

\counterwithin{equation}{section}
\VerbatimFootnotes

\newcommand{\alg}[1]{\mathfrak{#1}}

\begin{document}
\selectlanguage{english}
\title{{\color{Blue}\textbf{One-Point Functions of Non-protected Operators in the $SO(5)$ Symmetric D3-D7 dCFT}}\\[12pt]} \date{}
%
%\begin{comment}
\author{\textbf{Marius de Leeuw},$^1$ \textbf{Charlotte Kristjansen},$^1$ \textbf{and Georgios Linardopoulos}$^{2, 3}$\footnote{E-mails: \href{mailto:deleeuwm@nbi.ku.dk}{deleeuwm@nbi.ku.dk}, \href{mailto:kristjan@nbi.ku.dk}{kristjan@nbi.ku.dk}, \href{mailto:glinard@inp.demokritos.gr}{glinard@inp.demokritos.gr}.}
\footnote{Published in Journal of Physics A: Mathematical and Theoretical ``A Memorial Volume for Petr P. Kulish".}
\\[12pt]
$^1$ Niels Bohr Institute, University of Copenhagen, \\
Blegdamsvej 17, 2100 Copenhagen $\emptyset$, Denmark \\[6pt]
$^2$ Institute of Nuclear and Particle Physics, N.C.S.R., ``Demokritos",\\
153 10 Agia Paraskevi, Greece.\\[6pt]
$^3$ Department of Physics, National and Kapodistrian University of Athens,\\
Zografou Campus, 157 84 Athens, Greece.}
%\end{comment}
%
\maketitle
\begin{abstract}
\normalsize{\noindent
We study tree level one-point functions of non-protected scalar operators in the defect CFT, based on ${\cal N}=4$ SYM, which is
 dual to the $SO(5)$ symmetric D3-D7 probe brane system with non-vanishing instanton number. Whereas symmetries prevent operators from the $SU(2)$ and $SU(3)$ sub-sectors from having non-vanishing one-point functions, more general scalar conformal operators, which in particular constitute Bethe eigenstates of the integrable $SO(6)$ spin chain, are allowed to have non-trivial one-point functions.
For a series of operators with a small number of excitations we find closed expressions in terms of Bethe roots
for these one-point functions, valid for any value of the instanton number. In addition, we present some numerical results for operators with more excitations. }
\end{abstract}
%\newpage
%\tableofcontents \normalsize
\newpage
\section[Introduction and Motivation]{Introduction and Motivation \label{Section:IntroductionMotivation}}
Probe D3-D7 as well as probe D3-D5 brane systems allow for a configuration of the probe where the resulting dual
field theory has a co-dimension one defect carrying 2+1 dimensional Poincare symmetry and hosting 2+1 dimensional
Dirac fermions. Whereas the D3-D5 brane set-up
conserves half of the supersymmetries of $AdS_5\times S^5$, the D3-D7 set-up does not conserve
any supersymmetry. For this reason the latter set-up has been viewed as a promising arena for studying strongly coupled
fermion systems of relevance for condensed matter physics~\cite{Rey08}.

\indent The D3-D7 probe brane system describing ${\cal N}=4$ SYM with a co-dimension one defect comes in two variants, one where the geometry of the probe
brane is $AdS_4\times S^4$ and one where it is $AdS_4\times S^2\times S^2$. In the D3-D5 brane case the geometry
of the probe brane is $AdS_4\times S^2$. Unlike the D3-D5 brane case, the D3-D7 brane set-up is unstable to fluctuations
of the embedding coordinates of the spherical part of the brane geometry~\cite{MyersWapler08}, meaning that the mass of
a certain associated fluctuation mode violates the BF bound~\cite{BreitenlohnerFreedman82a}. One way of dealing
with this instability consists in embedding the probe D7 brane in the full black D3 brane metric instead of in its near horizon limit, $AdS_5\times S^5$~\cite{DavisKrausShah08}. This approach was followed in~\cite{DavisKrausShah08}, where the probe D3-D7 brane system
was used to model the transition between Hall plateaux and in~\cite{KristjansenSemenoff16} where the same system
was argued to behave as a fractional topological insulator. Another regularization method consists in introducing a cut-off
in the $AdS$ direction in the $AdS_5\times S^5$ background and taking this cut-off to infinity while sending the mass
of the unstable mode to the BF bound~\cite{KutasovLinParnachev11}, a method which was used
in~\cite{KutasovLinParnachev11} in a study of
transitions between conformal and non-conformal behaviour of 2+1 dimensional fermions and
 in~\cite{MezzaliraParnachev15} to further address the transition between
quantum Hall plateaux. A third way in which the D7 brane embedding can be stabilized is by introducing a sufficiently large background
gauge field flux or an instanton number on the spherical part of the D7 brane geometry. This strategy was suggested for the $SO(5)$ symmetric probe brane
geometry $AdS_4\times S^4$ in~\cite{MyersWapler08} and generalized to the $AdS_4\times S^2\times S^2$ case
in~\cite{BergmanJokelaLifschytzLippert10}. In the former case the background gauge field forms an instanton bundle on the
$S^4$ with a certain associated instanton number and in the second case the gauge field gives rise to a magnetic monopole
flux on the two two-spheres. In a variant of this approach only one two-sphere in the $AdS_4\times S^2 \times S^2$ geometry
gets a magnetic monopole flux. The corresponding probe D7 brane can be viewed as a state
consisting of a number of blown up D5 branes and is accordingly denoted as a giant D5 brane. Giant D5-branes have an interpretation as Hall states~\cite{KristjansenSemenoff12,KristjansenPourhasanSemenoff13,HutchinsonKristjansenSemenoff14}.

When the D3-D7 brane set-up is stabilized through the addition of a flux or an instanton number, the probe brane embedding is, strictly speaking, described by a so-called fuzzy funnel solution~\cite{ConstableMyersTafjord99}, however, the non-commutative nature of the embedding coordinates is usually ignored when the probe brane models are studied with the condensed matter perspective in mind.
In the present case we shall study the fuzzy funnel solution corresponding to the $SO(5)$ symmetric D7 brane embedding
 from the dual field theory perspective. The field theory in question is a non-supersymmetric defect conformal field theory
 with symmetry group $SO(2,3)\times SO(5)$.
 It consists of ${\cal N}=4$ SYM in a 3+1 dimensional bulk coupled
 to a flat 2+1 dimensional defect separating two regions of space-time with different ranks of the gauge group.
 The difference in the rank of the gauge group comes about because the string theory statement that the background gauge field has an instanton number $d_G$ on $S^4$ is equivalent to to the statement that $d_G$ of the $N$ D3-branes terminate on the D7-brane.
 In the field theory language,
 the difference in the rank of the gauge group is implemented by five of the six scalar fields of ${\cal N}=4$ SYM acquiring
 non-vanishing and space-time dependent vacuum expectation values (vevs) on one side of the defect.
 The simplest observables of defect conformal field theories are one-point functions which in the present set-up, due to the vevs, can be non-vanishing already at tree level.
 Conformal single trace scalar operators of ${\cal N}=4$ SYM constitute a closed $SO(6)$ sector at one-loop and can be characterized as being eigenstates of the integrable $SO(6)$ spin
chain with the eigenvalue being equal to the conformal dimension. The focus of the present paper will be on one-point functions
of such operators. In a closely related set-up consisting
 of a D3-D5 brane system with flux the study of tree-level one-point functions revealed interesting connections to
 integrability~\cite{deLeeuwKristjansenZarembo15,Buhl-MortensenLeeuwKristjansenZarembo15,deLeeuwKristjansenMori16} and allowed for a test of the AdS/dCFT set-up at the classical level~\cite{NagasakiYamaguchi12}. What is more, a study of one-loop one-point functions lead to a positive test of the AdS/dCFT set-up at the quantum level~\cite{Buhl-MortensenLeeuwIpsenKristjansenWilhelm16a,Buhl-MortensenLeeuwIpsenKristjansenWilhelm16c}.
% in a situation where both conformal symmetry and supersymmetry are partially broken~\cite{}.
In the present
 D3-D7 brane set-up so far only tree-level one-point functions of chiral primaries have been calculated~\cite{KristjansenSemenoffYoung12b}. Here, we
 extend the study to the more interesting case of non-protected operators which, in particular, is needed if one wants to explore possible relations to integrability. In this connection, let us mention that integrability in relation to two-point functions and anomalous dimensions of operators has been studied for other gauge gravity set-ups involving D7-branes,
including a D7-D3 probe-brane system where the D7-brane geometry is
$AdS_5\times S^3$~\cite{Erler:2005nr,Correa:2008av,MacKay:2010zb} and a
D7-O7-D3 set-up which involves an orientifold plane~\cite{Berenstein:2002zw,Stefanski:2003qr,Chen:2004mu,Chen:2004yf}.

 The present investigations are reported as follows. We start by reviewing in section~\ref{sec:Fuzzy} how the $SO(5)$ symmetric fuzzy funnel solution is realized in the field theory language. Next, in section~\ref{sec:spinchain} we recall a few facts about the integrable $SO(6)$ spin chain.
Section~\ref{sec:Onepoint} is devoted to the evaluation of one-point functions. First, we argue that operators belonging to the $SU(2)$ and $SU(3)$ sub-sectors of ${\cal N}=4$ SYM have vanishing tree-level one-point functions due to symmetries.\footnote{Notice that
the $SU(3)$ sector like the $SO(6)$ sector is only closed to one-loop order.}
Subsequently, we derive closed expressions for the one-point function of the
BMN vacuum as well as a number of excited states with few excitations and finally we list a number of numerical results for operators with more excitations. Section~\ref{sec:Conclusion} contains a discussion and conclusion.

\section{The fuzzy funnel solution\label{sec:Fuzzy}}

The probe D3-D7 brane system we will consider consists of a single D7 brane embedded in the usual $AdS_5\times S^5$ background, generated by $N$ D3-branes, in such a way that it wraps an $S^4$ inside $S^5$ and an $AdS_4$ inside $AdS_5$. Moreover, a background gauge field forms an instanton bundle on the $S^4$ with a certain instanton number\cite{MyersWapler08}.
The table below shows the relative orientation of the branes in flat space.
\\
\\

\setlength{\intextsep}{1pt plus 1.0pt minus 2.0pt}
\renewcommand{\arraystretch}{1.1}
\begin{table}[h]\begin{center}\begin{tabular}{|c||c|c|c|c|c|c|c|c|c|c|}
\hline
& $t$ & $x_1$ & $x_2$ & $x_3$ & $x_4$ & $x_5$ & $x_6$ & $x_7$ & $x_8$ & $x_9$ \\ \hline
\text{D3} & $\bullet$ & $\bullet$ & $\bullet$ & $\bullet$ &&&&&& \\ \hline
\text{D7} & $\bullet$ & $\bullet$ & $\bullet$ & & $\bullet$ & $\bullet$ & $\bullet$ & $\bullet$ & $\bullet$ & \\ \hline
\end{tabular}\caption{The D3-D7 system.}\end{center}\end{table}

In the field theory language the D7 brane gives rise to a co-dimension one defect which we assume to be located at $x_3=z=0$.
The defect supports a hypermultiplet of defect fields which interact among themselves and with the fields of ${\cal N}=4$
SYM which
live in the bulk. The field theory living on the defect has not been worked out in detail, but will not play any role for the following
analysis.

 The fuzzy funnel solution of the probe brane system maps onto to an $\mathfrak{so}(5)$
 symmetric solution of the classical equations
 of motion for the scalar fields of ${\cal N}=4$ SYM~\cite{ConstableMyersTafjord01a}
\begin{eqnarray}
\frac{d^2\Phi_i^{\text{cl}}}{dz^2} = \left[\Phi_j^{\text{cl}}, \left[\Phi_j^{\text{cl}}, \Phi_i^{\text{cl}}\right]\right], \qquad i = 1,\ldots,6.
\end{eqnarray}
A solution of these equations with the appropriate symmetry was found by \cite{CastelinoLeeTaylor97, ConstableMyersTafjord01a}:
\begin{eqnarray}
\Phi_i ^{\text{cl}}= \frac{G_i \oplus 0_{N-d_G}}{\sqrt{8}\,z}, \quad i = 1,\ldots,5, \qquad \Phi_6^{\text{cl}} = 0, \hspace{0.5cm} z>0. \label{Solution1}
\end{eqnarray}
where the $G_i$ are matrices whose commutators generate a $d_G$-dimensional irreducible representation of $\mathfrak{so}(5)$. Below we explain how to construct these matrices starting from the four-dimensional gamma matrices. For $z<0$
the classical fields (which are matrices of size $(N-d_G) \times (N-d_G)$) are vanishing. The rank of the gauge group is hence (broken) $SU(N)$ for $z>0$ and
$SU(N-d_G)$ for $z<0$.

\paragraph{Gamma matrices}Let $\sigma_i$ be the Pauli matrices, then we introduce the following four-dimensional gamma matrices
\begin{align}\label{eq:Gamma}
&\gamma_{i=1,2,3} =
\begin{pmatrix}
0 & -i \sigma_i\\
i\sigma_i & 0
\end{pmatrix},
&\gamma_4 =
\begin{pmatrix}
 0 & 1 \\
 1 & 0
\end{pmatrix},
&&\gamma_5 =
\begin{pmatrix}
1 & 0 \\
0 & -1
\end{pmatrix}.
\end{align}
It is readily checked that they satisfy the Clifford algebra
\begin{align}
\{ \gamma_i,\gamma_j\} = 2 \delta_{ij}.
\end{align}
The commutators of these gamma matrices $\gamma_{ij} = [\gamma_i,\gamma_j]/2$ form the four-dimensional spin representation of $\alg{so}(5)$.

\paragraph{Properties} Let us discuss a few useful properties of the gamma matrices. First, for any set of signs $s_i = \pm$ such that $s_1s_2s_3s_4s_5 = 1$ there exist similarity transformations $U_s$ such that
\begin{align}
U_s \gamma_i U^{-1}_s = s_i \gamma_i. \label{signs}
\end{align}
Moreover, let $P\in S_5$ be any permutation of $\{1,2,3,4,5\}$, then there exist a similarity transformation $U_P$ such that
\begin{align}
U_P \gamma_i U^{-1}_P = -\text{sign}(P) \gamma_{P_i}.
\end{align}
In other words, we have the freedom to relabel our gamma matrices.

\paragraph{G-matrices} We can now construct higher dimensional representations of $\mathfrak{so}(5)$
 by considering the coproduct. Define
\begin{align}
\Delta^{(n)} \gamma_i = \underbracket{\gamma_i \otimes 1 \otimes\ldots \otimes 1}_n + 1 \otimes \gamma_i \otimes 1 \otimes\ldots \otimes 1 + \ldots .
\end{align}
Clearly the matrices $\Delta^{(n)} \gamma_i $ no longer satisfy the Clifford algebra, but since $\Delta^{(n)}$ commutes with the commutator $\Delta^{(n)}\gamma_{ij}$ still forms a representation of $\alg{so}(5)$. However, this representation is reducible.

Consider the completely symmetric orthonormal basis vectors $v_i$ of $\otimes^n\mathbb{C}^4$, \textit{i.e.}\ $v_i$ is invariant under any permutation of the $n$ copies of $\mathbb{C}^4$. It is easy to show that the number of such vectors is equal to
\begin{equation}
d_G = \frac{1}{6}(n+1)(n+2)(n+3). \label{dG}
\end{equation}
Let us construct the projector
\begin{align}
&\Pi: \mathbb{C}^{4n} \rightarrow \mathbb{C}^{d_G} &&\Pi = (v_1, v_2, \ldots, v_{d_G})^t,
\end{align}
where $t$ stands for transposition. Since $\Pi$ projects onto the symmetric subspace, it has the following properties
\begin{align}
&\Pi \, \Pi^t = 1_{d_G},
&& [\Pi^t \Pi, \Delta^{(n)} A ]= 0,
&& [\Pi^t \Pi,A\otimes A\otimes\ldots ]= 0,
\end{align}
for any $4\times 4$ dimensional matrix $A$. We then finally define the $G$ matrices as the symmetrized version of $\Delta^{(n)} \gamma_i $
\begin{align}
G_i = \Pi \Delta^{(n)} \gamma_i \Pi^t.
\end{align}
The matrices $G_{ij}\equiv[G_i,G_j]/2$ form a $d_G$-dimensional representation of $\alg{so}(5)$.

\paragraph{Properties} All the similarity transformations that we discussed above for the gamma matrices trivially extend to the $G$-matrices as well by taking
\begin{align}
U \rightarrow \Pi\, (U\otimes U\otimes\ldots ) \Pi^t.
\end{align}
Moreover we have the following Casimir $G_i G_i = n(n+4)\mathbb{I}$.

\section{The integrable $SO(6)$ spin chain \label{sec:spinchain}}

It is well-known \cite{MinahanZarembo03} that the mixing of the single trace scalar operators at one-loop in $\mathcal{N} = 4$ SYM, is described by an integrable $\mathfrak{so}\left(6\right)$ spin chain with Hamiltonian
\begin{align}
\mathbb{H} = \frac{\lambda}{8\pi^2}\sum_{j = 1}^L \left(\mathbb{I}_{j,j + 1} - \mathbb{P}_{j,j + 1} + \frac{1}{2} \, \mathbb{K}_{j,j + 1}\right), \qquad \label{SpinChainHamiltonian}
\end{align}
where $\mathbb{I}$, $\mathbb{P}$ and $\mathbb{K} $
%
%\begin{align}
%&\mathbb{I} \cdot |\Phi_a\Phi_b\rangle = |\Phi_a\Phi_b \rangle,
%&&\mathbb{P} \cdot |\Phi_a\Phi_b \rangle = |\Phi_b\Phi_a \rangle ,
%&&\mathbb{K} \cdot |\Phi_a\Phi_b \rangle = \delta_{ab} \, \sum_{c = 1}^6|\Phi_c\Phi_c \rangle,
%\end{align}
%
are the identity, permutation and trace operators respectively. The spin chain eigenstates correspond to the conformal operators
and the eigenvalues to the associated conformal dimensions. Highest weight eigenstates for an $\mathfrak{so}\left(6\right)$ spin chain of length $L$ are characterized by
three sets of Bethe roots
\begin{equation}
\{u_{1,j}\}_{j=1}^{N_1}, \hspace{0.5cm}\{u_{2,j}\}_{j=1}^{N_2}, \hspace{0.5cm}\{u_{3,j}\}_{j=1}^{N_3},
\end{equation}
where
\begin{equation}
0\leq N_1\leq L, \hspace{0.5cm} 0\leq N_2\leq N_1/2, \hspace{0.5cm} 0\leq N_3\leq N_2,
\end{equation}
and the corresponding $\mathfrak{so}\left(6\right)$ representation is $(J_1,J_2,J_3)=(L-N_1,N_1-N_2-N_3, N_2-N_3)$ with
$J_1\geq J_2\geq J_3\geq 0$.
A translation to the language of fields will be given in the subsequent section.
The roots have to fulfill the Bethe equations
\begin{eqnarray}
\left(\frac{u_{1,i}+i/2}{u_{1,i}-i/2}
\right)^L&=&\prod_{j\neq i}^{N_1}\frac{u_{1,i}-u_{1,j}+i}{u_{1,i}-u_{1,j}-i}\prod_{k=1}^{N_2}\frac{u_{1,i}-u_{2,k}-i/2}{u_{1,i}-u_{2,k}+i/2}
\prod_{l=1}^{N_3}\frac{u_{1,i}-u_{3,l}-i/2}{u_{1,i}-u_{3,l}+i/2}, \\
1&= & \prod_{l\neq i}^{N_2}\frac{u_{2,i}-u_{2,l}+i}{u_{2,i}-u_{2,l}-i}
\prod_{k=1}^{N_1}\frac{u_{2,i}-u_{1,k}-i/2}{u_{2,i}-u_{1,k}+i/2}, \\
1&=&\prod_{l\neq i}^{N_3}\frac{u_{3,i}-u_{3,l}+i}{u_{3,i}-u_{3,l}-i}
\prod_{k=1}^{N_1}\frac{u_{3,i}-u_{1,k}-i/2}{u_{3,i}-u_{1,k}+i/2}.
\end{eqnarray}
Furthermore, in order that the spin chain eigenstate correctly reflects the cyclicity property of the corresponding single trace operator the
momentum carrying roots $\{u_{1,j}\}_{j=1}^{N_1}$ have to obey the following relation.
\begin{equation}
\prod_{i=1}^{N_1}\frac{u_{1,i}+i/2}{u_{1,i}-i/2}=1. \label{momentum}
\end{equation}

\section{One-point functions --- general considerations\label{sec:Onepoint}}

One point functions in a dCFT are constrained by conformal symmetry to take the form
\begin{equation}
\langle{\cal O}_{\Delta}(x)\rangle=\frac{C_{\Delta}}{z^{\Delta}},
\end{equation}
where $\Delta$ is the conformal dimension in the theory without the defect and $z$ is the distance to the defect.
Our aim is to compute one-point functions of local, gauge invariant operators at tree level which involves replacing
all fields in an operator with their classical vevs, i.e.\ making the substitution
\begin{align}
{\cal O}=\mathcal{O}^{i_1\ldots i_L} \mathrm{tr}(\Phi_{i_1}\ldots\Phi_{i_L}) \rightarrow
\mathcal{O}^{i_1\ldots i_L} \mathrm{tr}(G_{i_1}\ldots G_{i_L}).
\end{align}
Following~\cite{deLeeuwKristjansenZarembo15,Buhl-MortensenLeeuwKristjansenZarembo15} we can implement the above
substitution by calculating the inner product $\langle \mathrm{MPS}|\mathcal{O}\rangle$ between the operator $\mathcal{O}$ and a matrix product state, which is defined as
\begin{align}
|\mathrm{MPS}\rangle_L = \mathrm{tr} \prod_{l=1}^L \left( |\Phi_i\rangle_l \otimes G_i \right). \label{MatrixProductState}
\end{align}
%
%At this point, we can make some general remark regarding one-point functions in our theory.
%
%First, by extending the similarity transformation $U_s$ to the $G$-matrices, we see that only states where all the flavors of %scalar fields occur an even number of times have a non-trivial overlap with the Matrix Product State.
The normalized one-point function is then given by
\begin{equation}
\langle {\cal O}\rangle= \frac{\langle \mathrm{MPS}| {\cal O}\rangle}{\sqrt{\langle {\cal O}|{\cal O}\rangle} }.
\end{equation}
In the following we will work in the basis of complex fields given by
\begin{align}
&W = \Phi_1 + i \Phi_2 ,
&&Y = \Phi_3 + i \Phi_4 ,
&&Z = \Phi_5 + i \Phi_6 \\
&\bar{W} = \Phi_1 - i \Phi_2 ,
&&\bar{Y} = \Phi_3 - i \Phi_4 ,
&&\bar{Z} = \Phi_5 - i \Phi_6.
\end{align}
From our similarity transformation $U_P$ we see that the specific assignment of which complex fields are comprised of which $\Phi$'s is irrelevant.

 Let us now translate from
 the roots of an eigenstate to the field content of the corresponding operator. We
denote the highest weight as $\vec{q}=(1,0,0)$ and the three simple roots of $\mathfrak{so}\left(6\right)$ as $\vec{\alpha}_1=(1,-1,0)$,
$\vec{\alpha}_2=(0,1,-1)$, and $\vec{\alpha}_3=(0,1,1)$. Then a solution of the Bethe equations for a chain of
length $L$ and with given
values of $N_1, N_2$ and $N_3$ corresponds to the
highest weight state of the representation
$\vec{w}=L\vec{q}-N_1\vec{\alpha}_1 - N_2\vec{\alpha}_2 - N_3\vec{\alpha}_3=(L-N_1,N_1-N_2-N_3,N_2-N_3)$. We
associate weights
with the six complex fields in the following way
\begin{align}
& Z\sim \vec{q}, && W\sim \vec{q}-\vec{\alpha}_1, && Y\sim \vec{q}-\vec{\alpha}_1-\vec{\alpha}_2, \\
& \bar{Z}\sim \vec{q}-2\vec{\alpha}_1-\vec{\alpha}_2-\vec{\alpha}_3, && \bar{W} \sim  \vec{q}-\vec{\alpha}_1-\vec{\alpha}_2-\vec{\alpha}_3, && \bar{Y} \sim  \vec{q}-\vec{\alpha}_1-\vec{\alpha}_3.
\end{align}
This means that we map the vacuum state of the spin chain to the operator $|0\rangle=\mathrm{tr} Z^L$.
A level one root then creates an excitation of type $W$, a level two root transforms an $W$-excitation to a
$Y$-excitation and a level three root introduces a barred field.

It is easy to show that there is a similarity transformation $U_{a,b}$ that transforms the $G$-matrices as
\begin{align}
&U_{a,b} (G_1 \pm i G_2) U^{-1}_{a,b} = a^{\pm1} (G_1 \pm i G_2),
&&U_{a,b} (G_3 \pm i G_4) U^{-1}_{a,b} = b^{\pm1} (G_3 \pm i G_4),
&&U_{a,b} G_5 U^{-1}_{a,b} = G_5, \nonumber
\end{align}
for any complex numbers $a,b$. From this, it directly follows that the only non-trivial one-point functions are built from operators that have $\# W = \# \bar{W}$ and $\# Y = \# \bar{Y}$. Expressed in terms of Bethe roots this means that $N_1=N_2+N_3$
and $N_2=N_3$. This in particular means that $N_1$ has to be even. We thus conclude that only operators of the following type can have
non-vanishing one-point functions
\begin{align}
(L,N_1,N_2,N_3)= (L,M,M/2,M/2), \hspace{0.5cm} M = \mbox{even}.
\end{align}

Following the same strategy as for the $SU(2)$ and $SU(3)$ spin
chains~\cite{deLeeuwKristjansenZarembo15,deLeeuwKristjansenMori16}
one can show that the third conserved charge of the present integrable spin chain, i.e.\ the $SO(6)$ chain, annihilates the
present matrix product state, i.e.\
\begin{equation}
Q_3\cdot |\mbox{MPS}\rangle=0.
\end{equation}
For details we refer to appendix~\ref{Appendix:Charge3}.
As in the previously studied cases this implies that the Bethe roots at the first level have to consist of pairs of roots with opposite signs. Repeating the analysis
of~\cite{deLeeuwKristjansenMori16} one finds that the roots at the second and third level have to
fulfill that $\{u_{2,i}\}=\{-u_{2,i}\}$ and $\{u_{3,i}\}=\{-u_{3,i}\}$. Hence, if the number of roots at the second level is even
the roots have to come in pairs with opposite signs and if the number is odd, one of the roots has to be zero and the
rest have to be paired. Similarly for roots at the third level.

\paragraph{SU(2) and SU(3) sectors} The SO(6) spin chain contains two well-known subsystems, namely the SU(2) and the SU(3) integrable spin chains for which the spin chain states correspond to operators built from respectively two and three
holomorphic fields. These sub-chains, however, do not contain any barred fields and by the general arguments outlined in the previous section we find that operators from the corresponding subsectors have vanishing one-point functions.

\section{One-point functions of specific operators}

\paragraph{Vacuum:}
To determine the one-point function of the BMN vacuum $|0\rangle=\text{tr}Z^L$ we have to evaluate
\begin{align}
\langle\text{MPS}|0\rangle = \mathrm{tr}\,G_5 ^{L}.
\end{align}
The general form of the matrices $G_5$ is the following\footnote{Note that we are rotating all the G-matrices by a similarity transform (the same for each value of $n$) so that the diagonal of $G_5$ has the ordering \eqref{G5matrices}. This in particular implies that $G_{5(n=1)}=-\gamma_5$.}
\begin{align}
G_5 = 2\,\Big\{\underset{(n+1) \ \text{terms}}{\underbrace{\left\{-\frac{n}{2}, \ldots\right\}}}, \ \underset{2n \ \text{terms}}{\underbrace{\left\{-\frac{n}{2} + 1, \ldots\right\}}}, \ \ldots, \ \underset{j \cdot (n-j+2) \ \text{terms}}{\underbrace{\left\{-\frac{n}{2}+j-1, \ldots\right\}}}, \ \ldots, \ \underset{2n \ \text{terms}}{\underbrace{\left\{\frac{n}{2} - 1, \ldots\right\}}}, \ \underset{(n+1) \ \text{terms}}{\underbrace{\left\{\frac{n}{2}, \ldots\right\}}}\Big\}, \label{G5matrices}
\end{align}
where $j = 1,2,\ldots,n+1$. The overlap therefore becomes
\begin{align}
\left\langle\text{MPS}|0\right\rangle = \text{tr}\, G_5 ^L = \sum_{j=1}^{n+1} \left[ j\left(n-j+2\right)\left(-n+2j-2\right)^L\right] \equiv V_n\left(L\right). \label{VacuumOverlap1}
\end{align}
This sum is readily worked out in terms of Bernoulli polynomials to give
\begin{align}
\left\langle\text{MPS}|0\right\rangle = \left\{\begin{array}{l} 0, \quad L \ \text{odd} \\[12pt] 2^L \cdot \left[\frac{2}{L+3} \, B_{L+3}\left(-\frac{n}{2}\right) - \frac{\left(n+2\right)^2}{2\left(L+1\right)} \, B_{L+1}\left(-\frac{n}{2}\right)\right], \quad L \ \text{even}. \end{array}\right.
\end{align}
In the large-$n$ limit we find:
\begin{align}
\langle\text{MPS}|0\rangle \sim \frac{n^{L+3}}{2\left(L+1\right)\left(L+3\right)} + O\left(n^{L+2}\right), \qquad n \rightarrow \infty.
\end{align}
Normalizing the one-point function we find
\begin{align}
\frac{\langle\text{MPS}|0\rangle}{\sqrt{\langle 0|0\rangle}}= \frac{1}{2\left(L+1\right)\left(L+3\right)} \frac{1}{\sqrt{L}} \Big[n^{L+3} + O\left(n^{L+2}\right)\Big]\cdot\frac{1}{z^L}, \qquad n \rightarrow \infty, \label{Correlator4}
\end{align}
where here and in the following we leave out a factor of $(\pi^2/\lambda)^{L/2}$ originating from field theory propagators
in combination with the factor $\sqrt{8}$ in the definition of the vevs, cf.\ eqn.~(\ref{Solution1}).
%In~\cite{KristjansenSemenoffYoung12b} it was shown
%in full agreement with the large-$n$ limit of the one-point function \eqref{Correlator2} that was found in %\cite{KristjansenSemenoffYoung12b}.

\paragraph{Chiral primaries:}
Tree level one point functions of chiral primaries were calculated in~\cite{KristjansenSemenoffYoung12b} by exploiting the
fact that these operators are in a one-to-one correspondence with spherical harmonics on $S^5$ and picking out those
particular harmonics which are $SO(5)$ symmetric. In particular, it was found that there existed only one such chiral primary
for a given even value of $L$. The chiral primary considered in the previous paragraph is not identical to the one of the same length considered in~\cite{KristjansenSemenoffYoung12b} but has a non-vanishing projection on the latter.

\paragraph{Konishi operator:} The simplest non-protected state is the Konishi operator
\begin{align}
|\mathcal{K}\rangle = \mathrm{tr}\, \Phi_i\Phi_i.
\end{align}
For this state we find
\begin{align}
\langle\mathrm{MPS}|\mathcal{K}\rangle = \frac{1}{48}n(n+1)(n+2)(n+3)(n+4),
\end{align}
which follows immediately from the Casimir relation
\begin{align}
\mathrm{tr}\, G_iG_i = n(n+4) d_G.
\end{align}

\paragraph{States $(L,2,1,1)$:}

These states obviously include the Konishi operator just considered and need to have vanishing auxiliary roots and paired momentum carrying roots. As a consequence, the eigenstates of the dilatation operator take the simple form
\begin{align}
|p\rangle :=\sum_{n_1<n_2} ( e^{i p (n_1-n_2)} + e^{i p (n_2-n_1+1)} ) | \ldots\mathcal{X}_{n_1}\ldots \bar{\mathcal{X}}_{n_2}\ldots\rangle
- 2\sum_{n_1} ( 1+ e^{ip} ) | \ldots \bar{Z}_{n_1}\ldots\rangle,
\end{align}
where $(\mathcal{X},\bar{\mathcal{X}}) =(Y,\bar{Y}),(\bar{Y},Y),(W,\bar{W}),(\bar{W},W)$. Moreover, the momentum $p$ satisfies the Bethe equation
\begin{align}
\left[\frac{u+\frac{i}{2}}{u-\frac{i}{2}} \right]^{L+1}= e^{ip(L+1)} =1.
\end{align}
The solution is given by
\begin{align}
u = \frac{1}{2} \cot \frac{2m\pi}{L+1},
\end{align}
for $m=1,\ldots,L$. The norm of the state is readily computed to be
\begin{align}
\langle p | p \rangle = 4L(L+1).
\end{align}
For $n=1$, the $G$-matrices are the gamma matrices \eqref {eq:Gamma} which all anti-commute. Thus, the overlap with the matrix product state can be computed exactly by summing geometric series. In particular
\begin{align}
\langle \mathrm{MPS}|p\rangle := 32\sum_{n_1<n_2} (-1)^{n_1-n_2}( e^{i p (n_1-n_2)} + e^{i p (n_2-n_1+1)} ) - 8\sum_{n_1} (1+ e^{ip} ),
\end{align}
since $\mathrm{tr} [ \gamma_5^{L-2} \gamma_i\gamma_j ] = 4\delta_{ij}$.
A straightforward computation then shows
\begin{align}
\langle\mathcal{O}_{L,2,1,1}\rangle_{n=1} = \frac{|\langle \textrm{MPS}|p\rangle|}{\sqrt{\langle p | p \rangle}} = 8\sqrt{\frac{L}{L+1}}\frac{u^2-\frac{1}{2}}{u^2+\frac{1}{4}}\sqrt{\frac{u^2+\frac{1}{4}}{u^2}}.
\end{align}
For general $n$ and even $L$, we can prove the following formula for the trace of the product of $G$-matrices
\begin{eqnarray}
\lefteqn{
\mathrm{tr} \big[ G_5^{L-x-2} (G_a+iG_b)G_5^x(G_a-iG_b)\big] = }\nonumber\\
&& \hspace{1.5cm}\frac{1}{2}\sum_{m = 0}^{\left\lfloor x/2 \right\rfloor} {x \choose 2m} \cdot 2^{2m} \cdot \bigg\{n\left(n+4\right)V_n\left(L-2m-2\right) \nonumber
 - V_n\left(L-2m\right)\bigg\} \\
&&\hspace*{1.5cm}- 2\sum_{m = 0}^{\left\lfloor (x-1)/2 \right\rfloor}{x \choose 2m+1} \cdot 2^{2m+1} \cdot V_n\left(L-2m-2\right), \quad \label{Gformula}
\end{eqnarray}
where $V_n$ is given by \eqref{VacuumOverlap1}. The formula \eqref{Gformula} is valid for all pairs of $a\neq b=1,2,3,4$. Then, one can generalize the one-point function to the general $n$-dimensional representation. We find that the one-point function is given by
\begin{align}
\frac{\langle\mathcal{O}_{L,2,1,1}\rangle}{\langle\mathcal{O}_{L,2,1,1}\rangle_{n=1}} =
\frac{u^2}{u^2-\frac{1}{2}}\sum_{j}^n j^L \frac{(n+2)^2-j^2}{8} \frac{[u^2+\frac{(n+2)j+1}{4}][u^2-\frac{(n+2)j-1}{4}]}{[u^2+(\frac{j+1}{2})^2][u^2+(\frac{j-1}{2})^2]},
\end{align}
where the sum runs over the even/odd integers depending on whether $n$ is even or odd. This structure resembles the structure that was found for higher dimensional representations for the D3-D5 system \cite{Buhl-MortensenLeeuwKristjansenZarembo15}.

\paragraph{Operators with $M>2$:} For $M > 2$, it is in general not possible to solve the Bethe equations analytically.
In appendix~\ref{Appendix:One-PointFunctions} we list a number of results on one-point functions for all Bethe states of
length $L\leq 6$ that have a non-trivial one-point function. We compute the one-point function for $n=1,2,3,4$ where $n$ is related to the dimension of the representation for the vevs as in equation~(\ref{dG}). For completeness we also list the corresponding Bethe roots. The Bethe wave functions, which
are essential for the computation of the one-point function, are obtained by explicitly diagonalizing the Hamiltonian
and picking out the eigenstates that are highest weight. The prospects of improving this procedure and of obtaining
a closed formula for the one-point functions is discussed below.

\section{Discussion and Conclusion\label{sec:Conclusion}}

With the present paper we have taken a first step towards the calculation of one-point functions of non-protected operators
in the dCFT theory dual to the $SO(5)$ symmetric D3-D7 probe brane system with non-trivial instanton number. Most importantly, we have derived a selection rule that determines which operators have non-trivial one-point functions. This selection rule in particular shows that all operators from the $SU(2)$ and $SU(3)$ subsectors of ${\cal N}=4$ SYM
(except for the BMN vacuum state) have vanishing one-point functions. In addition, we have calculated analytically all non-trivial one-point functions of operators with two excitations for any rank of the representation of the $SO(5)$ invariant vevs.

In the dCFT dual to the $SO(3)\times SO(3)$ symmetric probe D3-D5 brane system with flux the one-point functions
of the conformal non-protected operators of the $SU(2)$ sub-sector were non-vanishing and it was possible to express
all of these in a closed formula valid for any rank of the representation of the $SU(2)$ invariant vevs~\cite{deLeeuwKristjansenZarembo15,Buhl-MortensenLeeuwKristjansenZarembo15}. The formula in question
was furthermore expressed in terms of certain determinants of Bethe roots, the same determinants which occur in the norm formula for the Bethe eigenstates~\cite{Gaudin:1983,Korepin:1982}. A closed formula
of determinant form has likewise been found for tree-level one-point functions of operators from the $SU(3)$ sub-sector but
so far only in the case of the two-dimensional representation for the vevs~\cite{deLeeuwKristjansenMori16}.

A pressing question
is of course whether there exists a closed formula for the tree-level one-point functions in the present dCFT
with $SO(5)$ invariant vevs. As one-point functions of the $SU(2)$ and $SU(3)$ sub-sector of ${\cal N}=4$ SYM are trivial one
has to consider conformal operators of the full $SO(6)$ sector of the theory which correspond to Bethe eigenstates of the
integrable $SO(6)$ spin chain~\cite{MinahanZarembo03}. As a strategy for working towards a closed formula we
have generated a set of numerical data on the one-point functions which can be found in~\ref{Appendix:One-PointFunctions}.
The search for a closed formula is somewhat complicated by the fact that a closed formula for the norm of the Bethe eigenstates of the $SO(6)$ spin chain does not yet exist, except for a conjectured version defined via a recursive
relation~\cite{BissiGrignaniZayakin12}.
Furthermore, the very construction of Bethe eigenstates of the $SO(6)$ spin chain whose
Bethe equations involve three layers of nesting is somewhat demanding. It would be interesting and would facilitate the
endeavour of calculating one-point functions in the present set-up if the recently developed method to efficiently
construct the Bethe eigenstates of $SU(N)$ spin chains~\cite{GromovLevkovich-MaslyukSizov16} could be generalized to the $SO(N)$ case as well.

Inspired by~\cite{Nagasaki:2011ue,NagasakiYamaguchi12} it was suggested in~\cite{KristjansenSemenoffYoung12b} that one might be able to compare the
string and field theory observables in the present gauge-gravity set-up in a certain double scaling limit
\begin{equation}
\lambda \rightarrow \infty, \hspace{0.5cm} n\rightarrow \infty, \hspace{0.5cm}\frac{\lambda}{n^2}\,\,\,\,\,\, \mbox{fixed},
\end{equation}
where $n$ is related
to the rank of the representation carried by the vevs by eqn.~(\ref{dG}), the argument being that
$\frac{\lambda}{n^2}$ might be taken small even if both $\lambda$ and $n$ are large. This idea was shown to work
for tree-level one-point functions of chiral primaries where a match between field and string theory
results was indeed found~\cite{KristjansenSemenoffYoung12b}. A similar check of the idea in the
case of the D3-D5 brane set-up~\cite{NagasakiYamaguchi12} has recently been extended to the quantum level, i.e.\
to next to leading order in the double scaling parameter and an exact match was found between the one-point function
of the BMN vacuum~\cite{Buhl-MortensenLeeuwIpsenKristjansenWilhelm16a,Buhl-MortensenLeeuwIpsenKristjansenWilhelm16c} as well as the expectation
value of a straight Wilson line~\cite{deLeeuw:2016vgp} calculated in respectively gauge and string theory.
The D3-D5 set-up preserves half of the supersymmetries of the $AdS_5\times S^5 $ space~\cite{ConstableMyersTafjord99} and preserves the conformal
symmetry of the defect even at the quantum level~\cite{ErdmengerGuralnikKirsch02}. As opposed to this the present D3-D7 set-up does not conserve any supersymmetry~\cite{ConstableMyersTafjord01a}. It would be very interesting to extend the quantum computations of the D3-D5 set-up to the present one in order to investigate whether the quantum match for the former
hinges on supersymmetry still being partly present.

As mentioned in the introduction the D3-D7 probe brane system has found numerous applications as a model for condensed matter systems involving strongly coupled 2+1 dimensional fermions.
It would be interesting to investigate in more detail what implications the behaviour of the here studied one-point functions have for these systems.

\paragraph{Acknowledgements.}
We thank G. Semenoff and K.\ Zarembo for useful discussions. M.d.L.\ and C.K.\ were supported in part by FNU through grants number DFF-1323-00082 and DFF-4002-00037. G.L.\ would like to thank the Niels Bohr Institute, Jens Hoppe and Konstantinos Zoubos for hospitality and support while this work was in progress.

\appendix
\section[Action of the Third Charge]{Action of the Third Charge \label{Appendix:Charge3}}
\begin{figure}
\begin{center}
\includegraphics[scale=0.5]{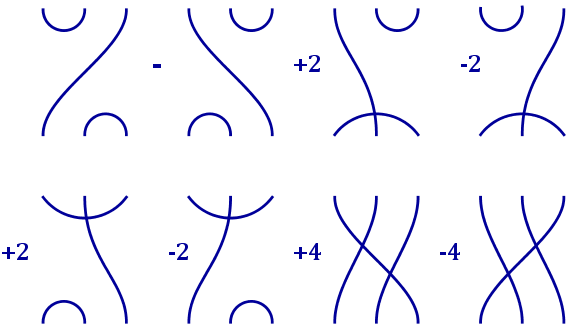}
\caption{The third charge $Q_3$.} \label{Graph:Charge3}
\end{center}
\end{figure}
Here we prove that the action of the third charge $Q_3$ on the matrix product state vanishes. The third charge is defined in terms of the Hamiltonian of the $\mathfrak{so}\left(6\right)$ spin chain \eqref{SpinChainHamiltonian} as
\begin{eqnarray}
Q_3 \equiv \sum_{j=1}^L Q_j, \qquad Q_j = \left[\mathbb{H}_{j-1,j},\mathbb{H}_{j,j+1}\right], \qquad \mathbb{H}_{j,j+1} \equiv \mathbb{I}_{j,j + 1} - \mathbb{P}_{j,j + 1} + \frac{1}{2} \, \mathbb{K}_{j,j + 1},
\end{eqnarray}
and $Q_j$ is depicted in figure \ref{Graph:Charge3}. For the MPS \eqref{MatrixProductState} we find:
\begin{eqnarray}
Q_j\cdot\text{MPS}_{(ijk)} &= \delta_{kj}G_s G_s G_i - \delta_{ij}G_k G_s G_s + 2\delta_{ik}G_j G_s G_s - 2\delta_{ik}G_s G_s G_j + 2\delta_{ij}G_s G_k G_s - \nonumber \\[6pt]
& \quad - 2\delta_{jk}G_s G_i G_s + 4G_k G_i G_j - 4G_j G_k G_i = \nonumber \\[6pt]
& = n\left(n+4\right)\left(\delta_{ij}G_k - \delta_{jk}G_i\right) + 8\left(G_{ij} G_k - G_i G_{jk}\right), \label{ActionCharge3}
\end{eqnarray}
which implies that $Q_3\cdot\left|\text{MPS}\right\rangle$ vanishes upon taking the trace over the entire MPS and summing the contributions \eqref{ActionCharge3} from all of its sites.
\section[Numerical results for $L\leq 6, n\leq 4$]{Numerical results for $L\leq 6, n\leq 4$ \label{Appendix:One-PointFunctions}}
This appendix contains the values of the one-point functions of the $SO(5)$ symmetric D3-D7 dCFT up to $n = 4$, $L = 6$. The one-loop eigenvalues are in units of $\lambda/8\pi^2$, so that the corresponding scaling dimensions are given by
\begin{align}
E = L + \frac{\lambda \gamma}{8\pi^2} + {\cal O}(\lambda^2).
\end{align}
Furthermore, $\gamma$ can be expressed in terms of the level one roots in the following way
\begin{equation}
\gamma=\sum_{i=1}^{N_1} \frac{1}{u_{1,i}^2+1/4}.
\end{equation}
The one-point functions are given in units of $\frac{1}{\sqrt{L}}(\frac{\pi^2}{\lambda})^{L/2}$.
\clearpage
\begin{table}[!t]
\begin{center}
\footnotesize\begin{eqnarray}
\begin{array}{|c|c|c|c|c|c|c|}
\hline &&&&&& \\
L & N_{1/2/3} & \text{eigenvalue } \gamma & \text{n=1} & \text{n=2} & \text{n=3} & \text{n=4} \\[6pt]
\hline  &&&&&& \\
{\color{blue}2} & {\color{red}2 \ 1 \ 1} & {\color{blue}6} & 20 \sqrt{\frac{2}{3}} & 40 \sqrt{6} & 140 \sqrt{6} & 1120 \sqrt{\frac{2}{3}} \\[6pt]
{\color{blue}4} & {\color{red}2 \ 1 \ 1} & {\color{blue}5+\sqrt{5}} & 20+\frac{44}{\sqrt{5}} & \frac{96}{5} \left(15+\sqrt{5}\right) & 84 \left(21-\sqrt{5}\right) & \frac{3584}{5} \left(10-\sqrt{5}\right) \\[6pt]
{\color{blue}4} & {\color{red}2 \ 1 \ 1} & {\color{blue}5-\sqrt{5}} & 20-\frac{44}{\sqrt{5}} & \frac{96}{5} \left(15-\sqrt{5}\right) & 84 \left(21+\sqrt{5}\right) & \frac{3584}{5} \left(10+\sqrt{5}\right) \\[6pt]
{\color{blue}4} & {\color{red}4 \ 2 \ 2} & {\color{blue}\frac{1}{2} \left(13+\sqrt{41}\right)} & 2 \sqrt{1410+\frac{25970}{3 \sqrt{41}}} & 16 \sqrt{3090+\frac{10710}{\sqrt{41}}} & 14 \sqrt{161490+\frac{140310}{\sqrt{41}}} & 896 \sqrt{690-\frac{670}{3 \sqrt{41}}} \\[6pt]
{\color{blue}4} & {\color{red}4 \ 2 \ 2} & {\color{blue}\frac{1}{2} \left(13-\sqrt{41}\right)} & 2 \sqrt{1410-\frac{25970}{3 \sqrt{41}}} & 16 \sqrt{3090-\frac{10710}{\sqrt{41}}} & 14 \sqrt{161490-\frac{140310}{\sqrt{41}}} & 896 \sqrt{690+\frac{670}{3 \sqrt{41}}} \\[6pt]
{\color{blue}6} & {\color{red}2 \ 1 \ 1} & {\color{blue}1.50604} & 3.57792 & 324.178 & 11338.3 & 98726 \\[6pt]
{\color{blue}6} & {\color{red}2 \ 1 \ 1} & {\color{blue}4.89008} & 9.90466 & 1724.55 & 19513.8 & 120347 \\[6pt]
{\color{blue}6} & {\color{red}2 \ 1 \ 1} & {\color{blue}7.60388} & 61.6252 & 1044.86 & 8830.95 & 49114.4 \\[6pt]
%{\color{blue}6} & {\color{red}4 \ 2 \ 2} & {\color{blue}8}       & 4.76832 & 2899.14 & 37483.7 & 247800 \\[6pt]
{\color{blue}6} & {\color{red}4 \ 2 \ 2} & {\color{blue}8}       & 3.41697 & 2077.52 & 26860.8 & 177573 \\[6pt]
{\color{blue}6} & {\color{red}4 \ 2 \ 2} & {\color{blue}2.26228} & 8.68876 & 1090.46 & 11963   & 166654 \\[6pt]
{\color{blue}6} & {\color{red}4 \ 2 \ 2} & {\color{blue}3.81374} & 13.8862 & 4479.21 & 43679.9 & 238186 \\[6pt]
{\color{blue}6} & {\color{red}4 \ 2 \ 2} & {\color{blue}5.33676} & 22.5105 & 2995.7  & 34577.8 & 216443 \\[6pt]
{\color{blue}6} & {\color{red}4 \ 2 \ 2} & {\color{blue}8.94875} & 78.0614 & 1813.66 & 16647.9 & 95264.6 \\[6pt]
{\color{blue}6} & {\color{red}4 \ 2 \ 2} & {\color{blue}10.1954} & 138.297 & 151.877 & 10250   & 80604.6 \\[6pt]
{\color{blue}6} & {\color{red}4 \ 2 \ 2} & {\color{blue}12.4431} & 369.992 & 4881.61 & 33331.2 & 159221 \\[6pt]
{\color{blue}6} & {\color{red}6 \ 3 \ 3} & {\color{blue}2.40409} & 10.796  & 3138.38 & 26455   & 403826 \\[6pt]
{\color{blue}6} & {\color{red}6 \ 3 \ 3} & {\color{blue}6.46525} & 24.1864 & 12630.2 & 158777  & 1.02938 \cdot 10^6 \\[6pt]
{\color{blue}6} & {\color{red}6 \ 3 \ 3} & {\color{blue}9.44931} & 16.2932 & 402.056 & 7995.5  & 64056.8 \\[6pt]
{\color{blue}6} & {\color{red}6 \ 3 \ 3} & {\color{blue}10.6753} & 248.809 & 3699.26 & 64674.8 & 456612 \\[6pt]
{\color{blue}6} & {\color{red}6 \ 3 \ 3} & {\color{blue}14.0061} & 774.89 & 13387.7 & 109348   & 589663 \\[6pt]
\hline
\end{array} \nonumber
\end{eqnarray} \normalsize \\
\caption{One-point functions of the $SO(5)$ symmetric D3-D7 dCFT up to $n = 4$, $L = 6$.}
\end{center}
%\end{table}
%
%\begin{table}[t]
\begin{center}
\footnotesize\begin{eqnarray}
\begin{array}{|c|c|c|c|c|c|}
\hline &&&&& \\
L & \text{eigenvalue } \gamma & u_{1,1} & u_{1,2} & u_{2,1} & u_{3,1} \\[6pt]
\hline &&&&& \\
{\color{blue}4} & {\color{blue}\verb|9.70156|} & \verb|0.700934|           & \verb|0.18859|            & \verb|0.519766|  & \verb|0.519766|  \\
{\color{blue}4} & {\color{blue}\verb|3.29844|} & \verb|0.307386+0.507862i| & \verb|0.307386-0.507862i| & \verb|0.608405i| & \verb|0.608405i| \\
{\color{blue}6} & {\color{blue}\verb|8|}       &\frac{\sqrt{3}}{{2}} & -\frac{1}{2\sqrt{3}} & \sqrt{\frac{1}{18}(1+\sqrt{37})} &
\sqrt{\frac{1}{18}(\sqrt{37}-1)}i
%& \verb|0.866025|
%& \verb|-0.288675|
%     & \verb|0.627285|  & \verb|0.53139i|
\\
{\color{blue}6} & {\color{blue}\verb|2.26228|} & \verb|0.753401+0.537295i| & \verb|0.753401-0.537295i| & \verb|0.740793i| & \verb|0.740793i| \\
{\color{blue}6} & {\color{blue}\verb|3.81374|} & \verb|0.228764+0.499841i| & \verb|0.228764-0.499841i| & \verb|0.592176i| & \verb|0.592176i| \\
{\color{blue}6} & {\color{blue}\verb|5.33676|} & \verb|0.499867|           & \verb|-1.11685|           & \verb|0.832238|  & \verb|0.832238|  \\
{\color{blue}6} & {\color{blue}\verb|8.94875|} & \verb|0.134791|           & \verb|1.0448|             & \verb|0.680098|  & \verb|0.680098|  \\
{\color{blue}6} & {\color{blue}\verb|10.1954|} & \verb|0.680815|           & \verb|0.143357|           & \verb|0.506298i| & \verb|0.506298i| \\
{\color{blue}6} & {\color{blue}\verb|12.4431|} & \verb|0.400681|           & \verb|0.118942|           & \verb|0.378798|  & \verb|0.378798| \\[6pt]
\hline
\end{array}
 \nonumber
\end{eqnarray} \normalsize \\
\caption{Bethe roots of the (L,4,2,2) states with $L=4,6$.}
\end{center}
\end{table}
\clearpage
\begin{table}[h!]
\begin{center}
\footnotesize\begin{eqnarray}
\begin{array}{|c|c|c|c|c|c|}
\hline &&&&& \\
\text{eigenvalue } \gamma & u_{1,1} & u_{1,2} & u_{1,3} & u_{2,1} & u_{3,1} \\[6pt]
\hline &&&&& \\
{\color{blue}\verb|2.40409|} & \verb|0.354396| & \verb|0.346921+0.975605i| & \verb|0.346921-0.975605i| & \verb|1.15779i|  & \verb|1.15779i| \\
{\color{blue}\verb|6.46525|} & \verb|0.666039| & \verb|0.341051+0.500039i| & \verb|0.341051-0.500039i| & \verb|0.235326i| & \verb|0.235326i| \\
{\color{blue}\verb|9.44931|} & \verb|0.498858| & \verb|0.0615056|          & \verb|1.03415i|           & \verb|0.894496i| & \verb|0.894496i| \\
{\color{blue}\verb|10.6753|} & \verb|0.141637| & \verb|0.455843+0.501973i| & \verb|0.455843-0.501973i| & \verb|0.488601i| & \verb|0.488601i| \\
{\color{blue}\verb|14.0061|} & \verb|0.952297| & \verb|0.418835|           & \verb|-0.11828|            & \verb|0.796688|  & \verb|0.796688| \\[6pt]
\hline
\end{array} \nonumber
\end{eqnarray} \normalsize \\
\caption{Bethe roots of the (L,6,3,3) states with $L = 6$.}
\end{center}
\end{table}
\bibliographystyle{JHEP}
\bibliography{Bibliography}
\end{document}